\documentclass{article}%
\usepackage{amsmath}
\usepackage{amsfonts}
\usepackage{amssymb}
\usepackage{graphicx}%
\begin{document}
\begin{flushright}
MZ-TH/10-37
\end{flushright}\begin{flushright}
October 2010
\end{flushright}

\medskip

\begin{center}
{\LARGE New Sum Rule Determination of the Nucleon\medskip\ }

{\LARGE Mass}\vspace{2cm}

{\large N.F. Nasrallah}$^{a,b}${\large \ and K. Schilcher}$^{b,c}$

$^{a}${\normalsize Lebanese University, Faculty of Science, Tripoli, Lebanon}

$^{b}${\normalsize Physics Department, American University of Beirut}

{\normalsize Beirut, Lebanon}

and

$^{c}${\normalsize Institut f\"{u}r Physik,
Johannes-Gutenberg-Universit\"{a}tD-55099 Mainz,
Germany\footnote{{\normalsize permanent address}}\vspace{2cm}}

\textbf{Abstract\medskip}
\end{center}

\begin{quote}
A new QCD calculation of the mass of the nucleon is presented. It makes use of
a polynomial kernel in the dispersion integrals tailored to practically
eliminate the contribution of the unknown $1/2^{+}$ and $1/2^{-}$ continuum.
This approach avoids the arbitrariness and other drawbacks attached to the
Borel kernel used in previous sum rules calculations. Our method yields stable
results for the nucleon mass and coupling for standard values of the
condensates. The prediction of the nucleon mass $m_{N}=(0.945\pm$ $.045)GeV$
is in good agreement with experiment.

\bigskip
\end{quote}

{\newpage}

\vspace*{2cm}

\section{Introduction{}}

The QCD sum rules method introduced by Shifman et al. \cite{1} has extended
the applicability of QCD far beyond simple perturbation theory. The method was
adapted to the case of nucleons by Ioffe \cite{2} and independently by Chung,
Dosch, Kremer and Schall \cite{3}. These authors showed how to approach one of
the fundamental problems of QCD, the calculation of baryon masses from the
Lagrangian and the vacuum condensates.

Nucleon mass sum rules start with the correlation function
\begin{equation}
\Pi(q)=i\int d^{4}xe^{iqx}\left\langle 0|\eta(x)\eta(0)|0\right\rangle
\label{6}%
\end{equation}
where $\eta$ is a nucleon interpolating field constructed from local QCD
operators with the quantum numbers of the nucleon. We will choose \cite{2}
\[
\eta=e^{abc}(u^{a}C\gamma^{\lambda}u^{b})\gamma_{5}\gamma^{\lambda}d^{c}\ .
\]
which couple maximally to the nucleon. The correlator can be decomposed in
terms of invariants,
\[
\Pi(q)=q_{\mu}\gamma^{\mu}\Pi_{1}(q^{2})+\Pi_{2}(q^{2})
\]
with $\gamma_{\mu}$ standing for the Dirac matrices. $\Pi(t=q^{2})$ is an
analytic function in the complex $t$-plane with a pole at $t=m_{N}^{2}$ and a
cut along the positive real axis starting at $t=(m_{N}+m_{\pi})^{2}$. The sum
rule methods can be traced back to the Cauchy formula
\begin{equation}
\frac{1}{2\pi i}\oint\Pi(t)P(t)\,dt=-\int_{0}^{R}\frac{1}{\pi}\text{Im}%
\Pi(t)P(t)\,dt \label{7}%
\end{equation}
where the kernel $P(t)$ is an arbitrary analytic function. The integral on the
left hand side (l.h.s.) is over a circle of radius $R$. If $R$ is taken large
enough, we can replace $\Pi(t)$\ on the l.h.s. by it's QCD and operator
product expansion (OPE) counterpart $\Pi_{QCD}(t).$ The right hand side
(r.h.s.) involves, apart from the nucleon pole, an integral over the cut,
consisting of a background plus a set of nucleonic resonances. Duality means,
that the OPE result on the l.h.s. of Eq. (\ref{7}), is equated to the hadronic
contribution on the right hand side. Traditionally the integrand on the r.h.s.
is approximated by the \textquotedblleft pole\ plus
continuum\textquotedblright\ model,
\begin{equation}
\frac{1}{\pi}\operatorname{Im}\Pi(q^{2})=\lambda_{N}^{2}\delta(q^{2}-m_{N}%
^{2})+\frac{1}{\pi}\theta(q^{2}-W^{2})\operatorname{Im}\Pi^{OPE}(q^{2})
\label{8}%
\end{equation}
Here $m_{N}$ is the position of the lowest lying pole with residue
$\lambda_{N}$, the coupling of the current to the nucleon state
\[
\left\langle 0|\eta|n\right\rangle =\lambda\Psi\ ,
\]
and an effective continuum threshold $W^{2}$ which is determined in the
calculation and on which the results depend sensitively.

Most sum rule studies of baryonic currents invoke a Borel transform of the
correlator, i.e. they use a kernel
\[
P(t)=e^{-t/M^{2}}%
\]
which introduces another, more or less arbitrary, parameter providing
exponential damping of the continuum(when it is small) and suppressing high
dimensional vacuum condensates(when it is large). Stability has to be
established under variations of the latter parameters.

The model of Eq.(\ref{8}) is rather unrealistic in cases with distinct higher
resonances. One may consider the resonances explicitly, e.g. for the nucleonic
correlator Eq.(\ref{6}) one can use in the sharp mass approximation
\begin{align*}
\frac{1}{\pi}\operatorname{Im}\Pi_{1}(q^{2})  &  =\lambda_{N}^{2}\text{
}\delta(q^{2}-m_{N}^{2})+\sum\limits_{i}\lambda_{i}^{2}\delta(q^{2}-m_{i}%
^{2})\\
\frac{1}{\pi}\operatorname{Im}\Pi_{2}(q^{2})  &  =m_{N}\lambda_{N}^{2}%
\delta(q^{2}-m_{N}^{2})+\sum m_{i}\lambda_{i}^{(+)2}\delta(q^{2}-m_{i}%
^{2})-\sum m_{i}\lambda_{i}^{(-)2}\delta(q^{2}-m_{i}^{2})
\end{align*}
For the nucleonic correlator considered here, there are four radial
recurrences $N^{+}(1440)$, $N^{-}(1535)$, $N^{-}(1650)$ and the $N^{+}(1710)$,
where $+$ or$-$ refers to the parity of the state \cite{PDG} and where the
couplings are unknown. The masses and widths of these states are well known
experimentally.It is seen from the above that the Borel sum rules involving
$\Pi_{2}(q^{2})$ are more reliable than the ones involving $\Pi_{1}(q^{2})$
because of the cancellation between resonances of opposite parities
\cite{JinTang}. With so many parameters to vary, the sum rule approach is
often viewed more as an art rather than a science.

To overcome these intrinsic ambiguities we have introduced some time ago a sum
rule method\ \cite{ACD}, originally called ACD, which exploits the analyticity
properties of the correlator to significantly reduce, in some cases
practically eliminate, the contribution of the continuum. The breakthrough in
the treatment of the continuum has been the introduction of an integration
kernel in the FESR tuned to suppress substantially the resonance energy region
above the ground state. This approach, specially adapted to eliminate
pronounced resonances, has been recently used to extract very precise values
of the light quark masses \cite{CAD09} and condensates \cite{Bordes}. Our
approach is based on the fact, that the contribution of the continuum in the
integral on the r.h.s. of Eq.(\ref{7}) arises mostly from the interval
\begin{equation}
2.0GeV^{2}\leq t\leq3.0GeV^{2} \label{11}%
\end{equation}
where the four nucleon resonances lie. This prompts us to choose
\begin{equation}
P(t)=\left(  1-\frac{t}{t_{0}}\right)  \text{ } \label{12}%
\end{equation}
where \ $t_{0}=2.52GeV^{2}$ is the location of the common midpoint of the
parity even and parity odd resonances. The value of $P(t)$ is small in
theregion of the resonances,moreover the contributions of the two parity even
resonances $N(1440)$ and $N(1770)$ (or the two parity odd ones $N(1535)$ and
$N(1650)$) come with opposite signs, so they tend to cancel in sum rules
involving both $\Pi_{1}(q^{2})$and $\Pi_{2}(q^{2})$. A residual model
dependency is still unavoidable as inelasticity, non-resonant background and
resonance interference are impossible to guess realistically. Also here our
approach helps, as a constant background is eliminated by the integration
kernel. Having thus minimized the contribution we will neglect it.

The theoretical side of the sum rule, in contrast, is in better shape. The
correlator (\ref{6}) is known including radiative corrections and OPE terms up
to dimension d = 9 \cite{2},\cite{Jamin},\cite{Sadovnikova}. There exists the
usual uncertainty about the precise values of the condensates and the validity
of the factorization assumption used for the higher dimensional condensates.
We note that the kernel Eq.(\ref{12}) will introduce only low dimension
condensates into the calculation which are known (or at least estimated).

Apart from the references cited above there are a few more attempts to
evaluate the sum rule (\ref{7}) involving a high sophistication on the
theoretical side which is not always commensurate with the primitive model
ansatz on the phenomenological side. We therefore think it necessary to
present for once an (almost) model independent investigation of the nucleonic
sum rule.

\section{The calculation}

The invariant amplitudes have poles at $t=m_{N}^{2}$%

\begin{align*}
\Pi_{1}(t)  &  =\frac{-\lambda_{N}^{2}}{(t-m_{N}^{2})}+...\\
\Pi_{2}(t)  &  =\frac{-m_{N}.\lambda_{N}^{2}}{(t-m_{N}^{2})}+...
\end{align*}

Having essentially eliminated the contribution of the continuum, it follows
then from Cauchy's theorem that

\bigskip%

\begin{align*}
|\lambda|^{2}P(m_{n}^{2})  &  =\frac{1}{2\pi i}\oint\limits_{|t|=R}\Pi
_{1}^{QCD}P(t)\\
m_{N}|\lambda|^{2}P(m_{n}^{2})  &  =\frac{1}{2\pi i}\oint\limits_{|s|=R}%
\Pi_{2}^{QCD}P(t)
\end{align*}
with%

\begin{align}
(2\pi)^{4}\Pi_{1}^{QCD}(t)  &  =A_{0}t^{2}\ln\frac{-t}{\mu^{2}}+A_{01}%
t^{2}\left(  \ln\frac{-t}{\mu^{2}}\right)  ^{2}+A_{4}\ln\frac{-t}{\mu^{2}%
}\nonumber\\
&  +A_{6}\frac{1}{t}++A_{61}\frac{1}{t}\ln\frac{-t}{\mu^{2}}+A_{8}\frac
{1}{t^{2}}+.... \label{3}%
\end{align}
and
\begin{equation}
(2\pi)^{4}\Pi_{2}^{QCD}(t)=B_{3}t\ln\frac{-t}{\mu^{2}}+B_{7}\frac{1}{t}%
+B_{9}\frac{1}{t^{2}}+... \label{4}%
\end{equation}

The coefficients $A_{i}$ and $B_{i}$ are defined as in \cite{Sadovnikova} but
for a factor $(2\pi)^{4}$and powers of $t$ taken explicitly%

\begin{align}
A_{0}  &  =-\frac{1}{4}(1+\frac{71}{12}a),A_{01}=\frac{a}{8}\nonumber\\
A_{4}  &  =-\frac{\pi^{2}}{2}\left\langle aGG\right\rangle \nonumber\\
A_{6}  &  =-\frac{2}{3}(2\pi)^{4}\left\langle \bar{q}q\bar{q}q\right\rangle
\left(  1-\frac{5}{6}a-\frac{1}{3}a\ln\frac{-t}{\mu^{2}}\right)  ,A_{61}%
=\frac{2}{9}(2\pi)^{4}\langle qqqq\rangle a\nonumber\\
A_{8}  &  =\frac{-1}{6}(2\pi)^{4}\mu_{0}^{2}\left\langle \bar{q}q\bar
{q}q\right\rangle \nonumber\\
B_{3}  &  =4\pi^{2}\left\langle \bar{q}q\right\rangle \left(  1+\frac{3}%
{2}a\right) \nonumber\\
B_{7}  &  =-\frac{4\pi^{4}}{3}\left\langle \bar{q}q\right\rangle \left\langle
aGG\right\rangle \nonumber\\
B_{9}  &  =(2\pi)^{6}\frac{136}{81}a\left\langle (\bar{q}q)^{3}\right\rangle
\nonumber
\end{align}
where $a=\frac{\alpha_{s}(\mu^{2})}{\pi}$. The terms $B_{7}$ and $B_{9}$ are
given in the factorization approximation. In $A_{8}$ we have taken
\[
\left\langle 0\right\vert \bar{q}q\bar{q}aG_{\mu\nu}^{c}\frac{\lambda_{c}}%
{2}\sigma^{\mu\nu}q\left\vert 0\right\rangle =\mu_{0}^{2}\left\langle \bar
{q}q\bar{q}q\right\rangle
\]
with the parameter $\mu_{0}^{2}=0.8GeV^{2}$ as advocated in \cite{Belyaev}. We
use the vacuum dominance approximation, also for the four-quark condensate,
$\left\langle \bar{q}q\bar{q}q\right\rangle \approx\left\langle \bar
{q}q\right\rangle ^{2}$. The question at which scale this relation holds is
resolved by allowing for generous errors. To avoid the double counting, we
keep the logarithmic $\ln(-t/\mu^{2})$ contribution in the polarization
operator but neglect its anomalous dimension. In any case anamalous dimesion
effects are very small \cite{Pivovarov}.

For the finite energy sum rule Eq.(\ref{7}), we need the known integrals of
the form
\[
I_{ik}=\frac{1}{2\pi i}\oint dtt^{i}\left(  \ln(-t)\right)  ^{k}%
\]
These are given in convenient form in \cite{Integrals}.

With our choice of P(t), we then get
\begin{align}
(2\pi)^{4}|\lambda|^{2}P(m_{N}^{2})  &  =-A_{0}R^{3}\left(  \frac{1}{3}%
-\frac{R}{4t_{0}}\right)  -A_{01}R^{3}(-\frac{2}{9}+\frac{2}{3}\ln(\frac
{R}{\mu^{2}})+\frac{1}{t_{0}}(\frac{R}{8}.-\frac{R}{2}\ln(\frac{R}{\mu^{2}%
}))\nonumber\\
&  -A_{4}R(1-\frac{R}{2t_{0}})-A_{6}-A_{61}(\ln\frac{R}{\mu^{2}}-\frac
{R}{t_{0}})+\frac{A_{8}}{t_{0}} \label{16}%
\end{align}%
\begin{equation}
(2\pi)^{4}|\lambda|^{2}m_{n}P(m_{N}^{2})=-B_{3}R^{2}\left(  \frac{1}{2}%
-\frac{R}{3t_{0}}\right)  -B_{7}+\frac{B_{9}}{t_{0}} \label{17}%
\end{equation}
In order to allow the cancellations introduced by $P(t)$ to take place, we
choose $R=2.92~GeV^{2}$, i.e. right after the fourth resonance. The nucleon
mass is obtained by dividing the two sum rules.

\section{Results and concusions}

With standard values of the condensates and using the factorization
approximation as discussed above, we get a value for the nucleon mass close to
the experimental value. For a more precise statement, we need do discuss
possible errors in the sum rule. It is important to distinguish two kinds of
errors, theoretical and experimental. A theoretical error of $\pm.03GeV$
arises from the uncertainty in the quark condensate. We used $\langle
qq\rangle=-(1.\,90\pm0.14)\times10^{-2}GeV^{3}$ at $\mu=2GeV$ corresponding to
the limits set by $m_{u}+m_{d}=(8.2\pm0.6)MeV$ \ (at scale $\mu=2GeV$)
\cite{CAD09} in the GMOR relation. Varying the scale parameter $\mu^{2}$
between $4GeV^{2}$ and $2GeV^{2}$ introduces an additional error of
$\pm0.03GeV$. Furthermore there is an error due to the strong coupling
constant. We use the latest comprehensive update analysis at the $\tau$-scale
\cite{Pich2010} which gives $\alpha_{s}(m_{\tau}^{2})=0.342\pm0.01$ and leads
to an error of $\pm0.6MeV$ in the nucleon mass. On other theory errors, like
the one due to the factorization assumption, we have no quantitative handle,
but under reasonable assumptions they turn out to be less important. We
estimate the experimental error, by varying $t_{0}$, the zero of the
polynomial $P(t)$,\ between \ $2.36GeV^{2}$ and $2.72GeV^{2}$. The resulting
error in the nucleon mass is
\[
m_{N}=(945\pm4.5)MeV
\]

In conclusion, we have presented a simple calculation of the nucleon mass
using a kernel in the dispersion integral tailored to minimize the
contribution of the unknown continuum without involving the higher order
unknown condensates. Using standard values of the known condensates, we obtain
for $m_{N}$ a value which agrees quite well with the experimental one.

\end{document}